\documentclass[%
 reprint,
 amsmath,amssymb,
 aps,
]{revtex4-1}

\usepackage{graphicx}
\usepackage{dcolumn}
\usepackage{bm}
\usepackage{multirow}
\usepackage{subcaption}
\usepackage{xfrac}
\usepackage{xcolor}
\usepackage{pgfplots}
\pgfplotsset{compat=1.14}
\tikzstyle{line}=[draw]

\newcommand{\Sec}{\tilde{S}^{\rm c}}
\newcommand{\Smc}{\tilde{S}}
\newcommand{\ceiling}{E} 
\newcommand{\ce}{\ceiling} 
\newcommand{\TT}{n_s}
\newcommand{\kR}{R^{\star}} 

\newcommand{\mce}{\mathcal{E}} 
\newcommand{\ord}{{\rm o}} 
\newcommand{\dis}{{\rm d}} 
\newcommand{\od}{{\rm c}} 
\newcommand{\as}{a_s}  
\newcommand{\rc}{r_c} 
\newcommand{\A}{X}  
\newcommand{\csh}{h_c}   
\newcommand{\drop}{droplet }
\newcommand{\Drop}{Droplet }
\newcommand{\m}{M}
\newcommand{\strip}{stripe }

\begin{document}
\preprint{APS/123-QED}
\title{Equilibrium Microcanonical Annealing for First-Order Phase Transitions}
\author{Nathan Rose}
\email{nathanrose68@gmail.com}
\affiliation{%
 Physics Department, University of Massachusetts, Amherst, Massachusetts 01003, USA\\
    1QB Information Technologies Inc., 458-550 Burrard Street, Vancouver, BC V6C 2B5, Canada
}%
\author{Jonathan Machta}%
 \email{machta@physics.umass.edu}
\affiliation{%
 Physics Department, University of Massachusetts, Amherst, Massachusetts 01003, USA\\
 Santa Fe Institute, 1399 Hyde Park Road, Santa Fe, New 
Mexico 87501 USA
}%

\date{\today}

\begin{abstract}
A framework is presented for carrying out simulations of  equilibrium systems in the microcanonical ensemble using annealing in an energy ceiling.  The framework encompasses an equilibrium version of  simulated annealing, population annealing and hybrid algorithms that interpolate between these extremes.  These equilibrium, microcanonical annealing algorithms are applied to the thermal first-order transition in the 20-state, two-dimensional Potts model. All of these algorithms are observed to perform well at the first-order transition though for the system sizes studied here, equilibrium simulated annealing is most efficient. 
\end{abstract}

\maketitle

\section{\label{sec:Intro}Introduction}
First-order phase transitions occur in a variety of systems and are characterized by a discontinuous first derivative in the free energy and phase coexistence at the transition temperature.  These characteristics pose difficulties for typical canonical ensemble Monte Carlo simulations because the transition rate between the coexisting phases decreases exponentially in system size at the transition temperature leading to exponential slowing. 

Exponential slowing can be substantially reduced using multi-canonical \cite{BeNe92} or Wang-Landau \cite{WaLa01,NoItWa11} methods.  A related means for reducing exponential slowing is to simulate the system in the microcanonical ensemble~\cite{MM07} rather than the canonical ensemble. For a temperature-driven, first-order phase transition, the coexisting phases will have distinct energies.  
Configurations with energies between the coexisting values of the energy possess domains of both phases separated by an interface and are very improbable in the canonical ensemble.  In the microcanonical ensemble, however, a droplet of a minority phase is stable at energies between the energies of the two coexisting phases, and so simulations can be carried out relatively efficiently at any energy in the coexistence region and exponential slowing is much reduced.  Nonetheless, in the microcanonical ensemble, there remains two or more barriers associated with condensation/evaporation \cite{BiChKo02,NoItWa11,ScZiJa16} transitions at the ends of the co-existence regions and, depending on the geometry of the system, also transitions in the topology of the growing droplet~\cite{LeZi90} so that exponential slowing cannot be entirely eliminated.

In this paper we introduce a class of microcanonical Monte Carlo annealing algorithms and show that they are effective for simulating the strong first-order transition in large-$q$ Potts models.  These algorithms are related to the well-known simulated annealing~\cite{KiGeVe83} algorithm but with two important differences.  First, annealing is carried out in energy rather than temperature, that is, we do microcanonical annealing.  Second, we introduce an additional resampling step in the annealing process in order to properly sample the equilibrium ensemble at every energy. 
One of the algorithms in this class that is studied here is a microcanonical version of population annealing~\cite{HuIb03,Mac10a,CaMa17}.

The purpose of annealing in the current setting differs from the usual objective of finding ground states or sampling thermal states for systems with rough free energy landscapes.  Here, microcanonical annealing is used to collect data at all energies in order to use reweighting to efficiently extract results at the transition temperature in the canonical ensemble.

The microcanonical ensemble is difficult to simulate using Markov chain Monte Carlo (MCMC) because of the constraint that proposed moves are accepted only if they do not change the energy. Here we circumvent that problem by simulating the ensemble of all equally weighted configurations with energies equal to {\em or below} a given energy ceiling, which we call the energy ceiling ensemble.  It is straightforward to transform results from the energy ceiling ensemble to the microcanonical ensemble.  An alternative approach to microcanonical Monte Carlo simulations is to introduce a kinetic energy term in the Hamiltonian and require that the total energy be fixed while the Monte Carlo moves change only the potential energy~\cite{MM07, ScZiJa16}.  This ``real'' microcanonical ensemble achieves the same objective of allowing a continuous range of potential energies for a fixed value of the energy control parameter. 

Simulated annealing \cite{KiGeVe83} consists of initializing a system at high temperature where equilibration using MCMC is easy, and gradually cooling the system following an annealing schedule that specifies a sequence of temperatures and number of MCMC sweeps at each temperature. The goal of simulated annealing is to find the ground state and, in its standard form, it is not a suitable algorithm for sampling equilibrium states.  
{\it Equilibrium} simulated annealing used here differs from standard simulated annealing in the choice of the starting configuration at the beginning of each annealing step.  In standard simulated annealing, the initial configuration at a new temperature is the final configuration generated by the MCMC at the previous temperature.  Note that this procedure is not well-defined in the energy ceiling ensemble since the last configuration generated by the MCMC may not be allowed at the new energy.   In equilibrium simulated annealing, the new configuration is randomly sampled with appropriate weights for the new annealing step from the set of configurations generated by the MCMC at the previous annealing step.  In the energy ceiling ensemble these weights are equal and the new state is randomly and uniformly chosen from the configurations generated at the previous energy that satisfy the new energy ceiling. Thus, to go from one annealing step to the next we ``resample'' (or, more correctly, ``subsample'') from the configurations generated at the previous step.

If multiple replicas of the system are simultaneously annealed, resampling can also be done among the set of replicas.  The resulting scheme is population annealing \cite{HuIb03,Mac10a,WaMaKa15,WaMaKa15b,AmMa18,BaMaWeHe18} and is an example of a sequential Monte Carlo \cite{DoFrGo01} algorithm. Population annealing in the canonical ensemble is an effective tool for simulating equilibrium systems with rough free energy landscapes, such as spin glasses~\cite{WaMaKa15b,WaMaMuKa17,AmMa18}.  Canonical population annealing has been applied to the first-order transition in Potts models~\cite{BaWeShJa17}. 
In microcanonical population annealing, resampling consists of discarding replicas that do not satisfy the new energy ceiling and making copies of those that do satisfy the new ceiling so that the population size remains constant. `Microcanonical' annealing in density rather than energy has been used to study hard sphere fluids~\cite{CaMa17}. Microcanonical population annealing is very similar to nested sampling~\cite{MaStWaFr14}.  Finally, we introduce a class of algorithms that interpolate between equilibrium simulated annealing and population annealing.  In these ``hybrid" annealing algorithms, multiple replicas are annealed and resampling is done from the union of time series of each replica. 

Annealing in the energy ceiling ensemble provides a direct measurement of the entropy, which can be used to calculate the canonical free energy and partition function, thus giving access to canonical ensemble observables at any temperature. The entropy or free energy can also be used to perform weighted averages over several simulation runs, reducing both statistical and systematic errors of observables in the microcanonical or canonical ensembles~\cite{Mac10a}. 

The primary goal of the paper is to introduce the equilibrium microcanonical annealing framework and test the performance of these algorithms for strong first-order transitions.  A secondary goal is to to obtain high precision results for the finite-size, two-dimensional, 20-state Potts model.
The paper is organized as follows.  In Sec.\ \ref{sec:Algorithm} we present the microcanonical annealing algorithms used in the simulations. In Sec.\ \ref{sec:model_and_observables} we define the $q$-state Potts model and the observables studied in the simulations. Section \ref{sec:simdet} gives details of the simulations and in Sec.\ \ref{sec:Results} we present  results from the simulations.  The paper closes with a discussion in Sec.\ \ref{sec:discussion}.

\section{\label{sec:Algorithm} Equilibrium Microcanonical Annealing}

In this section we present a general framework for equilibrium microcanonical annealing.  Two limiting cases of this framework are population annealing and equilibrium simulated annealing. In all cases we have an ``annealing schedule'' that consists of a sequence of energy ceilings $\{\ceiling^{(k)}\}$ and number of MCMC sweeps, $n_s(E^{(k)})$ that are carried out at each energy ceiling.  We study these algorithms in the context of the Potts model, which has a bounded, discrete set of energy levels.  For such systems the annealing schedule is naturally chosen as the set of energy levels  of the system and annealing is carried out from the highest energy level to the ground state. In the following discussion the parenthetical superscript `$(k)$' is suppressed.

Microcanonical annealing directly simulates the energy ceiling ensemble in which all configurations with energies less than or equal to the energy ceiling,  $\ceiling$ are equally probable.  In general, microcanonical annealing works with a population of replicas of the system that represent the energy ceiling ensemble. Each annealing step is carried out in two stages.  In the first stage a population of $R$ replicas of the system is updated using a MCMC updating procedure designed to equilibrate to the energy ceiling ensemble. During the course of this updating a pool of $\kR$ replicas of the system is saved where, generally, $\kR \geq R \geq 1$.  

The MCMC procedure used here is a single-spin-flip algorithm. The proposed update consists of flipping a randomly chosen spin.  Suppose that the initial spin state is $\alpha$ and the final spin state is $\gamma$ with energy $E_\gamma$.  The acceptance probability, $P(\alpha\rightarrow \gamma)$ of the proposed move is given by,
\begin{equation}
    P(\alpha\rightarrow \gamma)=
    \begin{cases} 
        1 & E_{\gamma} \leq \ceiling \\
        0 & E_{\gamma} > \ceiling . \\
    \end{cases}
    \label{eq:update}
\end{equation}
This MCMC algorithm is carried out for $\TT(\ceiling)$ sweeps on each replica, where a sweep consists of $N$ updates and $N$ is the number of spins in the system.  Clearly, this algorithm satisfies detailed balance with respect to the energy ceiling ensemble.  If, in addition, the algorithm is ergodic, that is, all configurations of the system can be reached from one another via a series of updates, then the algorithm will converge to the desired energy ceiling ensemble.  However, it is evident that this algorithm is {\it not} ergodic for all ceiling energies.  For example, for Ising-Potts models with $\ce$ sufficiently near the ground state, it is not possible to go from one ordered state to another and remain below the ceiling via single spin flip moves.  Nonetheless, full coverage of configuration space and accurate equilibrium results can be obtained either by using a population of replicas of the system as is done in population annealing and/or by combining multiple independent runs as is done in equilibrium simulated annealing.  

The second stage of microcanonical annealing consists of resampling with replacement from the pool of $\kR$ replicas with energy ceiling $\ceiling$ in order to generate a population of $R$ replicas that is approximately in equilibrium at the next lower value of the energy ceiling, $\ceiling^\prime$. Note that if $\kR$ is not sufficiently large the pool may contain no configurations with energy less than or equal to $\ceiling^\prime$, in which case the algorithm fails.  

The entire microcanonical annealing procedure is as follows:

\begin{enumerate}
\item A population of $R$ independent replicas of the system is initialized in equilibrium with the energy ceiling $\ceiling$, set to the highest energy level of the system.  
\item $\TT(\ce)$  sweeps of the MCMC procedure are performed on each member of the population and $\kR$ configurations generated from these MCMC sweeps are saved in the configuration pool. 
\item The energy ceiling is lowered to the next value $\ce^\prime$ in the annealing schedule ($\ceiling^\prime< \ceiling$) and $R$ configurations with energy less than or equal to $\ceiling^\prime$ are randomly drawn with replacement from the $\kR$ configurations in the pool.

\item Steps 2 and 3 are repeated until $\ceiling$ is equal to the ground state energy of the system.

\end{enumerate}

During the annealing step with ceiling energy $\ceiling$, the fraction $\epsilon(\ceiling)$ of replicas in the pool with energies greater than $\ceiling^\prime$ is called the ``culling fraction'' and is used to estimate entropies.

The equilibrium microcanonical annealing framework described above can be easily generalized to an equilibrium canonical annealing framework with the following two modifications.  First, the MCMC procedure must be chosen to equilibrate to a fixed temperature rather than a fixed energy ceiling.  Second, the resampling step must select from the configuration pool with Boltzmann reweighting, $e^{-(\beta^\prime - \beta)E_i}$, where $E_i$ is the energy of configuration $i$ and $\beta^\prime$ is the inverse temperature succeeding $\beta$ in the annealing schedule.  

\subsection{\label{sec:entropy}Entropy estimators}

There are two relevant entropies associated with microcanonical ensemble.  The energy ceiling or ``volume'' entropy, $S^c_\ceiling$ is defined as 
\begin{equation}   
    S^c_\ceiling = \log{\Sigma(\ceiling)},
\end{equation}
where $\Sigma(\ceiling)$ is the number of configurations of the system with energy less than or equal to $\ceiling$.  If $\ceiling'$ is the energy following $\ceiling$ in the annealing schedule (here the energy spectrum, $\mce$)  and if $\epsilon(\ceiling)$ is the culling fraction at energy $\ceiling$, then a recursive estimator for the ceiling entropy is
\begin{equation}
    \Sec_{\ceiling'} = \Sec_\ceiling + \log(1-\epsilon(\ce)),
    \label{eq:Sec}
\end{equation}
so that, 
\begin{equation}
    \Sec_{\ce} = \sum_{\ce_1 > \ce} \log(1-\epsilon(\ce_1)) + S^c_\infty,
    \label{eq:Secroll}
\end{equation}
where the summation is over all energies in the annealing schedule greater than $\ce$ and  $S^c_\infty$ is the logarithm of the total number of system states.
The standard definition of entropy in the microcanonical ensemble, $S_\ce$, sometimes referred to as the surface entropy, is given by 
\begin{equation}
\label{eq:Smc0}
\begin{aligned}
S_\ce &=\log[\Sigma(\ceiling) - \Sigma(\ceiling')],\\
& \approx  \log\left[\Sigma(\ceiling)\epsilon(\ceiling)\right],
\end{aligned}
\end{equation}
so that the estimator for this entropy is related to the ceiling entropy estimator by,
\begin{equation}
\label{eq:Smc}
 \Smc_\ceiling= \Sec_\ceiling + \log\epsilon(\ceiling).   
\end{equation}
In the thermodynamic limit the extensive parts of the two entropies are the same. In what follows we refer to $\Smc_\ceiling$ as the ``microcanonical entropy."

\subsection{\label{sec:level2}Estimators for microcanonical observables}

Suppose $A$ is an observable of the system such as the energy or magnetization.  An estimator $\tilde{A}^{\rm c}_\ceiling$ for the equilibrium value of $A$ in the energy ceiling ensemble is
\begin{equation}
\tilde{A}^{\rm c}_\ceiling = \frac{1}{\kR}\sum_{j=1}^{\kR} A^j
\end{equation}
where $A^j$ refers to the value of $A$ in the $j$th replica in the configuration pool for energy ceiling $\ceiling$.  We will generally use a superscript $c$ to indicate a quantity in the energy ceiling ensemble.

An estimator $\tilde{A}_\ce$ of the observable in the microcanonical ensemble at energy $\ceiling$ is obtained by averaging over only those replicas in the configuration pool that have energies at the ceiling, 
\begin{equation}
\tilde{A}_\ceiling = \frac{1}{\epsilon(\ce) \kR}\sum_{j=1}^{\kR} A^j \delta (E^j , \ceiling)
\end{equation}
where $\delta$ denotes Kronecker function and $E^j$ is the energy of replica $j$.  

\subsection{\label{sec:level3} Estimators for canonical observables}

The microcanonical entropy can be used to compute an observable in the canonical ensemble from the observable measured in microcanonical annealing.  An estimator of the canonical partition function at inverse temperature $\beta_c$ is given by
\begin{equation}
\label{eq:Z}
\tilde{Z}(\beta)=\sum_{\ce} e^{-\beta \ceiling + \Smc_\ceiling}
\end{equation}
where the sum is taken over all the energy levels of the system. The canonical energy distribution,
$\rho_\beta(\ceiling)$ is 
\begin{equation}
\rho_\beta(\ceiling)=\frac{e^{-\beta \ceiling + \Smc_\ceiling}}{\tilde{Z}(\beta)}.
\label{eq:canonical_histogram}
\end{equation} 
An estimator $\tilde{A}(\beta)$ of the observable $A$ in the canonical ensemble at inverse temperature $\beta$ is given by
\begin{equation}
\tilde{A}(\beta)=\sum_\ce \rho_\beta(\ce)\tilde{A}_{\ce},
\label{eq:MC_to_canonical}
\end{equation} 
and an estimator of the free energy is obtained from the definition,
\begin{equation}
\label{eq:F}
\beta \tilde{F} = - \log \tilde{Z}(\beta).
\end{equation}

For systems which undergo thermal first-order phase transitions, phase coexistence is manifested as a two-peaked canonical energy distribution $\rho_{\beta_c} (E)$ at the transition transition temperature, $\beta_c$.  In the thermodynamic limit, the two peaks are delta functions at the energies $E_\dis$ and $E_\ord$ of the coexisting disordered and ordered phases, respectively.  For an observable $A$, estimators of the coexisting  ordered and disordered values at the transition, $\tilde{A}_\ord$ and $\tilde{A}_\dis$ are, respectively,

\begin{subequations}
\begin{align}
\tilde{A}_\ord &= \frac{\sum_{\ceiling < \ce_\od} \rho_{\beta_c}(\ce)\tilde{A}_{\ce}}{\sum_{\ceiling < \ce_\od} \rho_{\beta_c}(\ce)} \\[5pt]
 \tilde{A}_\dis &= \frac{\sum_{\ceiling \geq \ce_\od} \rho_{\beta_c}(\ce)\tilde{A}_{\ce}}{\sum_{\ceiling \geq \ce_\od} \rho_{\beta_c}(\ce)},
\end{align}
\label{eq:peakObs}%
\end{subequations}
where the breakpoint energy, $\ce_\od$ must be chosen in the range $\ce_\ord < \ce_\od < \ce_\dis$. In the thermodynamic limit, the breakpoint energy can be chosen arbitrarily in this range.  For finite systems, an appropriate choice of $\ce_\od$ will help minimize finite-size corrections.  Clearly, $\ce_\od$ should be chosen in the middle of the range to avoid significant overlap with the broadened peaks in the energy distribution. 

\subsection{\label{sec:weight} Weighted averages}
It has been shown \cite{Mac10a,WaMaKa15b} that for population annealing in the canonical ensemble, multiple independent runs can be combined using weighted averaging to reduce both statistical and systematic errors.  In the case of the canonical ensemble the weight factor is proportional to the exponential of the free energy estimator for each run.  A similar result holds for equilibrium microcanonical annealing algorithms except that the weight factor is here proportional to the exponential of the entropy estimator \cite{CaMa17}.

Consider a collection of $M$ independent microcanonical annealing  runs, each with the same annealing schedule and the same population parameters $R$ and $\kR$.  Let $\tilde{A}_{\ceiling,m}$ denote the average value of an observable $A$ in the microcanonical ensemble from the $m$th run.  The best estimate $\bar{A}_\ce$ from the collection of $M$ runs is given by the weighted average, 
\begin{equation}
\bar{A}_\ce = \sum_{m=1}^{M}\omega_{\ce,m} \tilde{A}_{\ceiling,m} ,
\label{eq:reweighting}
\end{equation} 
where the weights are proportional to the exponential of the entropy estimator $\Smc_{\ceiling,m}$ of run $m$,
\begin{equation}
\label{eq:weight}
\omega_{\ce,m}=\frac{e^{\Smc_{\ceiling,m}}}{\sum_{m'=1}^{M} e^{\Smc_{\ceiling,m'}}}.
\end{equation}
An analogous result holds for the energy ceiling ensemble,  
\begin{equation}
\bar{A}_\ce^c = \sum_{m=1}^{M}\omega^c_{\ce,m}  \tilde{A}^c_{\ceiling,m} ,
\label{eq:reweightingc}
\end{equation}
where, 
\begin{equation}
\omega^c_{\ce,m}=\frac{e^{\Sec_{\ceiling,m}}}{\sum_{m'=1}^{M} e^{\Sec_{\ceiling,m'}}}.
\label{eq:weightc}
\end{equation}
For either ensemble, if a run fails at energy $\ce$ then for all energies $\ce^\prime \geq \ce$ the corresponding weight vanishes.

It is most straightforward to first derive the weighted average formula for the energy ceiling result and then the microcanonical result.  Consider the simple situation that at annealing step $\ce$ the weights for all runs are the same, $\omega^c_{\ce,m} \equiv 1/M$.  At the next annealing step with energy ceiling $\ce^\prime$, the population to be averaged over in run $m$ has a size $R\kR(1-\epsilon_{m}(\ce))$ where $\epsilon_{m}(\ce)$ is the culling fraction at energy ceiling $\ce$ in run $m$.  An unbiased average in the energy ceiling ensemble over many runs at energy ceiling $\ce^\prime$ should weight each run according to the size of its population, which requires that run $m$ be weighted by a factor proportional to $(1-\epsilon_{m}(\ce))$.  From Eq.\ \eqref{eq:Sec} we have that $(1-\epsilon_{m}(\ce))$ is the exponential of the volume entropy change from $\ce$ to $\ce^\prime$.  

In the more general case where runs already have differing weights at energy ceiling $\ce$ we have the recursion relation,
\begin{equation}
\omega^c_{\ce^\prime,m} \propto (1-\epsilon_m(\ce))\omega^c_{\ce,m},   
\end{equation}
where the constant of proportionality is independent of run and is set by the normalization of the weights.
From Eq.\ \eqref{eq:Sec} we see that $(1-\epsilon_{m}(\ce))$ is the exponential of the volume entropy change from $\ce$ to $\ce^\prime$.  Collapsing the  telescoping product of entropy changes yields the weight factor given in Eq.\ \eqref{eq:weightc}.

To obtain the analogous result for the microcanonical ensemble, note that the population that is averaged in the microcanonical ensemble is a factor $\epsilon_m(\ce)$ smaller than the population averaged in the ceiling ensemble.  Thus the microcanonical ensemble and ceiling ensemble weights are related by, 
\begin{equation}
    \omega_{\ce,m} \propto \epsilon_m(\ce)\omega^c_{\ce,m},
\end{equation}
and from Eq.\ \eqref{eq:Smc} we see that this transformation is precisely the transformation from $\Sec$ to $\Smc$, thus verifying \eqref{eq:weight}.

Formulas for the weighted average of the two entropies are slightly more complicated to derive because the entropy depends on a thermodynamic integration over all of the annealing steps, Eq.\ \eqref{eq:Secroll}.  Thus the weighted average volume entropy can be written as,
\begin{equation}
    \bar{S}^c_\ce = \log \prod_{\ce_1 > \ce} \, \sum_{m=1}^M (1 - \epsilon_{m}(\ce_1))\omega^c_{\ce_1,m} + \Sec_\infty.
\end{equation}
Expanding the definition \eqref{eq:weightc} of the weight factor and using Eq.\ \eqref{eq:Sec} to relate the culling factor to the change in entropy yields the telescoping product,
\begin{equation}
    \bar{S}^c_\ce = \log \prod_{\ce_1 > \ce} \frac{\sum_{m=1}^M \exp(\Sec_{\ce_1^\prime,m})}{\sum_{m=1}^M \exp(\Sec_{\ce_1,m})} + \Sec_\infty,
\end{equation}
where $\ce_1^\prime$ is the successor to $\ce_1$ in the annealing schedule.  After canceling terms in the numerator and denominator we have,
\begin{equation}
    \bar{S}^c_\ceiling = \log \frac{1}{M}\sum_{m=1}^{M}e^{\Sec_{\ceiling,m}},
    \label{eq:reweighting_S0}
\end{equation}   
and, analogously for the microcanonical entropy, we have
\begin{equation}
    \bar{S}_\ceiling = \log \frac{1}{M}\sum_{m=1}^{M}e^{\Smc_{\ceiling,m}}.
    \label{eq:reweighting_S}
\end{equation} 

We emphasize that for fixed $R$ and $\kR$,  results from weighted averaging are exact in the limit $M\rightarrow \infty$ for any fixed number of MC sweeps per run.  

Weighted averaging in the canonical ensemble is discussed in Refs.\ \cite{Mac10a,WaMaKa15b}.  Given $M$ independent simulations in the canonical ensemble, the best estimate $\bar{A}(\beta)$ of an observable $A$ in the canonical ensemble at inverse temperature $\beta$,  is given by
\begin{equation}
    \bar{A}(\beta)= \frac{ \sum_{m=1}^M \tilde{A}_m(\beta) e^{-\beta \tilde{F}_m}}{\sum_{m=1}^M e^{-\beta \tilde{F}_m}}
    \label{eq:free_energy_rw}
\end{equation} 
where $\beta \tilde{F}_m = -\log \tilde{Z}_m(\beta)$ is the free energy estimator of the $m$th run. In analogy to Eq.\ \eqref{eq:reweighting_S}, the best estimator of the free energy from multiple runs is given by,
\begin{equation}
-\beta \bar{F} = \log \frac{1}{M}\sum_{m=1}^{M}e^{-\beta \tilde{F}_m}.
\label{eq:reweighting_F}
\end{equation} 

Since our simulations directly produce results in the microcanonical ensemble, it is important to show that transforming from the microcanonical to the canonical ensemble commutes with carrying out a weighted average in either ensemble.  Specifically, we would like to show that, 
\begin{equation}
    \label{eq:commute_A}
    \bar{A}(\beta)=\frac{ \sum_{\ce}\bar{A}_\ce \ e^{-\beta \ce + \bar{S}_E} }{\sum_{\ce} e^{-\beta \ce + \bar{S}_\ce}},
\end{equation}
and
\begin{equation}
    \label{eq:commute_F}
    -\beta \bar{F}=\log \sum_{\ce} e^{-\beta \ce + \bar{S}_\ce}, 
\end{equation}
where the canonical weighted averages, $\bar{A}(\beta)$ and $\beta \bar{F}$ are defined in \eqref{eq:free_energy_rw} and \eqref{eq:reweighting_F}, respectively, and where on the RHS the barred quantities are microcanonical weighted averages defined in, Eqs.\ \eqref{eq:reweighting}, \eqref{eq:weight} and \eqref{eq:reweighting_S}.  To verify \eqref{eq:commute_F}, simply substitute the definition \eqref{eq:reweighting_S} of $\bar{S}_\ce$ on the RHS, exchange the order of summation and then invoke the definitions, \eqref{eq:Z} and \eqref{eq:F}. A similar but slightly more involved calculation verifies Eq.\ \eqref{eq:commute_A}.  Finally, we note that the weighted averages for  observables, $A_\ord$ and $A_\dis$, defined in Eqs.\ \eqref{eq:peakObs} are obtained from Eq.\ \eqref{eq:commute_A} with the summation over a restricted range of energies, $\ce<\ce_\od$ and $\ce \geq \ce_\od$, respectively. Alternatively, results analogous to Eq.\ \eqref{eq:free_energy_rw} for $A_\ord$ and $A_\dis$ are obtained by replacing $\tilde{F}$ by $\tilde{F}_\ord$ and $\tilde{F}_\dis$, respectivly, where these partial free energies are defined in the obvious way by limiting the range of summation in the partition function, Eq.\ \eqref{eq:Z}.

\subsection{Systematic errors}
\label{sec:errors}
It is possible to quantify systematic errors in microcanonical annealing algorithms from the run-to-run variance of the entropy estimator.  This result follows from the fact that weighted averages are exact in the limit $M \rightarrow \infty$.  Systematic errors are then given by the difference between weighted and unweighted averages in this limit. The result, derived for the case of population annealing ($R\gg 1$ and $\kR = R$) in Refs.\ \cite{WaMaKa15b,CaMa17}, is that the systematic error, $\Delta A_\ce$ in the estimator $\tilde{A}_\ce$ is given by 
the covariance of $\tilde{A}_\ce$ and $\Smc$,
\begin{equation}
    \label{eq:syserror}
    \Delta A_\ce =  {\rm cov}(\tilde{A}_\ce,\Smc) = {\rm var}(\Smc) \left[\frac{{\rm cov}(\tilde{A}_\ce,\Smc)}{{\rm var}(\Smc)}\right].
\end{equation}
The second, trivial identity in Eq.\ \eqref{eq:syserror} is useful because the ratio in the square brackets goes to a constant that depends on the observable $A$ as the simulation size, $\kR$ becomes large, while ${\rm var}(\Smc)$ is expected vanish as $1/\kR$ since the number of measurements going into estimating $\Smc$, also becomes large.  Note that the scaling of the simulation size depends on the type of annealing algorithm: in equilibrium simulated annealing $\kR$ becomes large with $R=1$ and in population annealing, $\kR=R$, while various options are possible for hybrid annealing.   In any case, systematic errors and their scaling with simulation size are characterized by the behavior of ${\rm var}(\Smc)$.  Although this result was originally derived for population annealing, it holds for the whole class of equilibrium microcanonical annealing protocols discussed here since it depends on only  the the validity of weighted averaging.  An analogous result holds for quantities in the energy ceiling ensemble.   

The systematic error of an observable in the canonical ensemble, $\Delta A(\beta)$ is given by,
\begin{equation}
    \label{eq:syserrorcan}
    \Delta A(\beta) =  {\rm cov}(\tilde{A}(\beta),\beta \tilde{F}) = {\rm var}(\beta \tilde{F}) \left[\frac{{\rm cov}(\tilde{A}(\beta),\beta \tilde{F})}{{\rm var}(\beta \tilde{F})}\right].
\end{equation}
After transforming from the microcanonical to the canonical ensemble, this result can be applied to quantify errors in canonical observables obtained from microcanonical simulations. Results for systematic errors for $A_\ord$ and $A_\dis$ can be obtained from the variance of the partial free energies $\tilde{F}_\ord$ by $\tilde{F}_\dis$, respectively, discussed at the end of Sec.\ \ref{sec:weight}.

The variance of entropy and free energy estimators are therefore a useful tool for comparing the performance of different versions of equilibrium microcanonical annealing.

\subsection{Relation to other annealing algorithms}

The case $R=1$ and $\kR \gg 1$ is a form of simulated annealing.  Simulated annealing is conventionally carried out in the canonical ensemble with the last configuration generated by the MCMC procedure used as the starting point for the next annealing step.  Thus, conventional simulated annealing is the case $R=\kR=1$.  If this conventional choice were applied in the microcanonical ensemble the algorithm would fail with probability $\epsilon$ at each annealing step and would be highly inefficient.  The case $R \gg 1$ and $\kR=R$ is microcanonical population annealing, which has previously been applied to study hard sphere systems \cite{CaMa17}.  We refer to the intermediate cases $1< R < \kR$ as hybrid annealing.  In what follows we abbreviate population annealing, simulated annealing and hybrid annealing as PA, SA and HA, respectively.

One of our goals is to understand the trade-offs related to the population size $R$.  One conclusion that seems evident is that a single large run of a hybrid algorithm with parameters $R=R_{\rm h}$ and $\kR=\kR_{\rm h}$ should outperform the weighted average of $M=R_{\rm h}$ runs of simulated annealing with $\kR=\kR_{\rm h}/M$ even though both simulations require nearly the same amount of computational work.  The argument is that resampling from the larger pool in the hybrid algorithm will be more effective than simply reweighting independent runs.  A related result was observed in comparing standard canonical simulated annealing with canonical population annealing for finding ground states of spin glasses~\cite{WaMaKa15}. 

\section{\label{sec:model_and_observables}Model and Observables}
\subsection{\label{sec:model}The $q$-state Potts Model}
The $q$-state Potts model \cite{Wu82} is a generalization of the Ising model consisting of $N$ interacting spins $\{s_i|i=1, \ldots, N\}$  each taking values $s_i \in \{1,...q\}$.  The energy, $\ce$ is given by,
\begin{equation}
    \ce=-\sum_{ \langle i,j \rangle} \delta(s_i,s_j), 
\end{equation}
where $\delta$ is the Kronecker delta function and the summation is over all the interacting pairs of spins.  For spins on a square (two-dimensional) lattice with nearest neighbor interactions and for $q>4$, the Potts model undergoes a thermal first-order phase transition at inverse temperature $\beta_c = \log(1+\sqrt{q})$.  Higher $q$-values result in stronger first-order phase transitions.  In our simulations we consider the two-dimensional, $20$-state Potts model on an $L\times L$ square lattice with periodic boundary conditions.      

The first-order phase transition of the Potts model in the canonical ensemble is between a disordered phase with magnetization zero and energy per spin $e_\dis$ and a magnetically ordered phase with magnetization per spin $m_\ord$ and energy per spin $e_\ord$.  Configurations with  energy per spin between these values possess an interface between ordered and disordered domains, leading to a free energy barrier proportional to the system size $L$.  

In the microcanonical ensemble configurations with an interface are stable for energies per spin between $e_\dis$ and $e_\ord$ making  it possible for an annealing algorithm to traverse the coexistence region nearly continuously.  Nonetheless, there is a series of four discontinuous transitions in the microcanonical ensemble, which introduce barriers between the disordered and ordered phases and increase equilibration times.  First, at a length scale dependent energy, $e_1(L)<e_\dis$ there is a transition corresponding to the condensation of an ordered droplet within a disordered background~\cite{BiChKo02,NoItWa11,ScZiJa16}.  As the system size increases, $e_1(L)$ approaches $e_\dis$.   As energy decreases, this droplet grows continuously until the system reaches energy $e_2$ where it can decrease its surface area by a discontinuous transition to a \strip configuration.  Decreasing the energy further results in the widening of the ordered \strip until at energy $e_3$ there is another discontinuous transition to a configuration with a disordered droplet in an ordered background~\cite{LeZi90}.  Finally, for $e_4(L)>e_\ord$, there is an evaporation transition where the disordered droplet vanishes, leaving the homogeneous ordered phase.  Mirroring $e_1(L)$, $e_4(L)$ depends on system size and approaches $e_\ord$ from above.

\subsection{Observables \label{sec:observables}}

In this section we define the observables measured in our simulations.  First, we measure the entropy using Eq.\ \eqref{eq:Smc} and use it to calculate the free energy and the canonical energy distribution of Eq.\ \eqref{eq:canonical_histogram} at the transition point $\beta_c$. From these quantities we calculate the following observables.

\paragraph*{Energies of ordered and disordered phase at coexistence}
 Using the canonical energy distribution and Eqs.\ \eqref{eq:peakObs} we measure the ordered and disordered phase energies per spin, $e_\ord$ and $e_\dis$, respectively.   

 \paragraph*{Peak ratio and disordered phase excess}
 We measure two observables related to the relative weights of the ordered and disordered peaks in the energy distribution at $\beta_c$, the peak ratio and the disordered phase excess, both of which are defined below.  
 
 The peak ratio $\rc$ at the transition temperature is defined as,
\begin{equation}
    \rc = \frac{\sum_{\ceiling < \ce_\od} \rho_{\beta_c}(\ce)} 
        {\sum_{\ce \geq \ce_\od} \rho_{\beta_c}(\ce)}.
    \label{peakRatio}
\end{equation}
The peak ratio is related to the difference in the free energies of the ordered and disordered phases, which in the thermodynamic limit are equal at the transition temperature.  However, since there are $q$ different ordered phases, the exact value of the peak ratio in the thermodynamic limit is $q$ at the transition temperature.

Since the peak ratio is not an observable of the form of Eq.\ \eqref{eq:MC_to_canonical}, the weighted average cannot be computed simply from Eq.\ \eqref{eq:free_energy_rw}.  We define another observable related to the peak ratio called the disordered phase excess $\A$, 
\begin{equation}
    \A = \sum_{\ceiling > \ce_\od} \rho_{\beta_c}(\ce) - \frac{1}{q + 1},
    \label{eq:disordered_excess}
\end{equation}
where the subtracted constant is chosen so the exact value of $\A$ in the thermodynamic limit is zero.  The disordered phase excess is used as a measure of accuracy of a simulation at the transition point. Unlike the peak ratio, a weighted average of $\A$ using Eq.\ \eqref{eq:free_energy_rw} gives the best estimate of $\A$ and is exact in the limit $M \rightarrow \infty$.

\paragraph*{Wrapping fraction} To estimate the locations $e_2$ and $e_3$ of the droplet/\strip transitions we define the wrapping fraction, $\omega_w$, as the average number of directions wrapped by connected paths of like spins. The number of wrapping directions takes values in $\left\{ 0,1,2\right\}$; in the disordered phase and the droplet state the wrapping number is zero, in the \strip state the wrapping number is one, while in the ordered phase the wrapping number is two.  

\paragraph*{\Drop fraction} In order to detect the evaporation transition between a configuration containing a disordered droplet and a homogeneous ordered phase, we consider the disordered cluster size histogram $\csh$, where here a disordered cluster is defined using the following procedure.  First, identify the largest connected cluster of adjacent spins having the same value.  Near the evaporation transition point this cluster will wrap both vertically and horizontally.  Label all spins in this cluster $``0"$ and label all other spins $``1"$.  Group adjacent spins labelled $``1"$ into clusters, and let  $\csh$ be the configuration averaged histogram of the sizes of these disordered clusters.

The disordered cluster size distribution is shown in Fig.\ \ref{fig:csh} for three energies. At energies lower than the evaporation transition point $e_4 (L)$, all clusters consist of excitations within a homogeneous ordered phase, and $\csh$ decays exponentially.  At the transition point $\csh$ is observed to be a power law distribution as the disordered droplet breaks down into smaller disordered regions. At energies greater than the transition point, a finite fraction of spins exist within a disordered droplet, and thus, in addition to a power law, $\csh$ contains a peak at some value greater than zero. 

In order to measure the \drop fraction $\omega_c$, which we define as the probability that a configuration contains a large droplet of disordered phase, we subtract the power law component from $\csh$ and integrate over the remaining peak.  In practice, this is done by finding the least squares fit of $\csh$ to a power law up to a cutoff size that is chosen so that the fit only covers the power law region of $\csh$.  What remains is a single peak centered on the characteristic droplet size for the given energy.  

When $\csh$ is averaged over many measurements, we expect the integral of this peak to be one for energies sufficiently above the transition point.  In other words, we expect every configuration to contain one droplet that is significantly larger than the droplet size distribution predicted by the power law.  As the energy approaches $e_4$ from above, this peak will decrease in size and $\omega_c$ will drop to zero as the power law behavior of $\csh$ crosses over to exponential decay.  In this regime our procedure becomes inaccurate as it becomes more difficult to fit $\csh$ to a power law, and in general this method of measuring $\omega_c$ is only valid near the transition point.  

\paragraph*{Magnetization integrated autocorrelation time}

The  magnetization per spin, $m$ for the Potts model is defined as,  
\begin{equation}
m= \frac{q(N_{\rm max}/N) - 1}{q-1},
\end{equation}
where $N_{\rm max}$ is the maximum number of spins taking the same value.
In the disordered phase $N_{\rm max}\approx N/q$ and the magnetization vanishes, while in the ordered phase $N_{\rm max} > N/q$.  As the temperature approaches zero, $N_{\rm max} \rightarrow N$ and the magnetization per spin approaches one.

The integrated autocorrelation time $\tau$ of the magnetization is defined by
\begin{equation}
    \tau = \frac{1}{2} + \frac{\sum_{t=1}^{\infty} ( \langle m(t_0)m(t_0 + t)\rangle-\langle m \rangle ^2)}{\langle m^2 \rangle -\langle m \rangle ^2}
    \label{eq:autocorrelation}
\end{equation} 
where time is measured in Monte Carlo sweeps and the brackets refer either to an ensemble average or an average over starting times $t_0$.  The integrated autocorrelation time sets the time between independent measurements.  Although the asymptotic decay of the correlation functions, i.e.\ the exponential autocorrelation time, is a better measure of the time to reach equilibrium, $\tau$ is usually also a good approximate measure of the equilibration time or, equivalently, the computational work required to equilibrate the system at a given energy.

The time series used to compute $\tau$ must be much larger than $\tau$ itself and and upper limit of the sum in Eq.\ \eqref{eq:autocorrelation} must be finite.  Here we use an iterative method described in \cite{OsSo04} that minimizes systematic errors and is used to set the upper limit of the sum defining $\tau$.

\section{Simulation details}
\label{sec:simdet}
We perform SA, HA, and PA simulations for the 2D 20-state Potts model for systems sizes $L = 30,40,50,$ and $60$.  For each size, the computational work, measured in sweeps, is kept the same for each algorithm.  In each case, we obtain results by using weighted averaging over $\m$ simulation runs.  The number of MC sweeps performed at each energy is equal to $\TT$ given by
\begin{equation}
    \TT(\ce)=
    \begin{cases}
        \as,& \text{if  } \ce > -\frac{N}{2}\\
        20 \as,& \text{if  } -\frac{N}{2} \geq \ce > -\frac{3N}{2}\\
        5 \as,& \text{if  } -\frac{3N}{2} > \ce 
    \end{cases}
    \label{eq:sweep_schedule}
\end{equation}
where $\as$ is a parameter that depends on the algorithm and is chosen to make the computational work roughly equal between algorithms for the same system size.   

This  {\it ad hoc} sweep schedule concentrates the majority of MC sweeps in the range of energies where the integrated autocorrelation time is highest.  We have found that this sweep schedule 
is an improvement compared to a uniform sweep schedule, but is not claimed to be optimal.

Simulation parameters for the three annealing algorithms are presented in Table \ref{table:simulation_parameters}.  Since both the number of proposed updates per MC sweep and the number of annealing steps in the simulation grow linearly with the system size $N$, the work per simulation grows quadratically for a given sweep parameter $\as$.  In order to keep the run times manageable, fewer MC sweeps ($\as$ decreased) are performed per annealing step for larger system sizes. In order to partially compensate for the decrease in the number of sweeps, reweighted averages are performed over more runs ($M$ increased) for larger system sizes.  

It is straightforward to take advantage of the massively parallel nature of the HA and PA algorithms by performing MC sweeps on independent replicas simultaneously. As the majority of computational work in these simulations is spent performing MC sweeps and only a small fraction is spent during the resampling step, significant improvement in wall clock time can be gained for PA and HA through parallelization despite doing the resampling sequentially.  

For PA the speedup due to parallelization is most dramatic, because the number of independent replicas is on the order of $10^5$.  Population annealing simulations were implemented on a single NVIDIA Tesla C2075 GPU using the $CUDA$ library using the methods described in Ref.\ \cite{BaWeBo17}.  One thread was assigned to each replica so that the total number of threads was equal to the population size.  Despite the simplicity of this implementation, we still observed a notable speed-up with PA when compared to SA with the same number of sweeps running on a CPU.  For instance, for our $L=50$ simulations, the PA runs were about $85$ times faster than SA runs.

For our HA simulations the population size of $100$ was not large enough to benefit from using a GPU though a significant speedup over SA was obtained using OpenMP with $16$ threads.

	\begin{table}
		\begin{tabular}{ |c|c|c|c|c|c|} 
			\hline
			Algorithm & $R$ & $L$ & $\as$ & $\kR/R$ &$M$\\
			\hline
            \multirow{4}{*}{SA} & \multirow{4}{*}{1}    & 30 &	$1.2 \times 10^6$ & $1.2 \times 10^6$ &  30	\\   	
                                &  					    & 40 &	$3.8 \times 10^5$ & $3.8 \times 10^5$ &  40 \\ 
                                &  					    & 50 &	$1.6 \times 10^5$ & $1.6 \times 10^5$ &  50	  \\
                                &  					    & 60 &	$7.5 \times 10^4$ &	$7.5 \times 10^4$ &  60	  \\
            \hline
            
            \multirow{4}{*}{HA}  & \multirow{4}{*}{100} & 30 &	$1.2 \times 10^4$ &$1.2 \times 10^3$&  30 \\ 
                                 & 					    & 40 &	$3.8 \times 10^3$ &$3.8 \times 10^2$&  40  \\ 
                                 &   					& 50 & $1.6 \times 10^3$ & $1.6 \times 10^2$ &  50 \\
                                 & 						& 60 & $7.5 \times 10^2$ & $7.5 \times 10$ &  60 \\

            \hline
            \multirow{4}{*}{PA}  & 	$8 \times 10^5$	    & 30 &	2	&   1  	&  30	  \\ 
                                 &	$2 \times 10^5$  	& 40 &	1.9	&   1	&  40	  \\ 
                                 &	$10^5$       		& 50 &  1.6& 1	&  50	  \\
                                 &	$7.5 \times 10^4$   & 60 &	1	&   1	&  60	  \\
            \hline

		\end{tabular}
		\caption{Simulation parameters for SA, HA, and PA. 
        Values in the table are system size $L$, population size $R$, the parameter $\as$ setting the number of MCMC sweeps (see Eq.\ \eqref{eq:sweep_schedule}), number of runs $M$ used for weighted averaging, and ratio of pool size to population size, $\kR/R$.
       }
\label{table:simulation_parameters}
\end{table}

\begin{figure}
    \includegraphics[width=\linewidth]{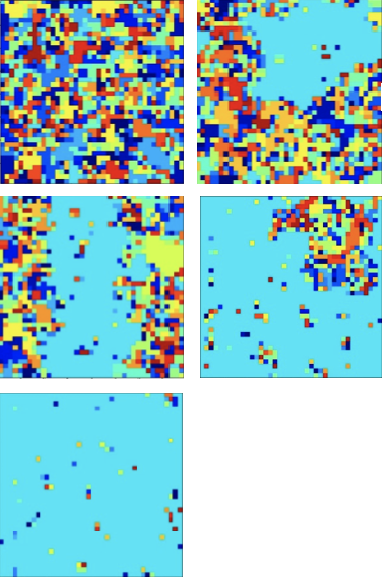}
    \caption{Equilibrium configurations of the 20-state Potts model with $L=40$ and periodic boundary conditions. From left to right and then bottom to top configurations are shown in the (1) homogeneous disordered phase, (2) disordered phase with an ordered droplet, (3) \strip phase, (4) ordered phase with disordered droplet, and (5) homogeneous ordered phase.}
    \label{fig:snapshots}
\end{figure}

\begin{figure}
    \hspace{-12pt}
    \begin{tikzpicture}[yscale=.9, xscale = .9]
        \pgfplotsset{every axis/.append style={
                        xlabel={$e$},
                        ylabel={$\rho(e)$},
                        ymin=10e-18,
                        ymax = 10e2
                        },
                        }
        \begin{axis}[ymode=log, legend pos= south east, execute at begin axis={
                    \draw (rel axis cs:0,0) -- (rel axis cs:1,0)
                          (rel axis cs:0,1) -- (rel axis cs:1,1);
                }]
        \addplot[red] table [mark=none, x=e, y=pe, col sep=space] {Plots/PA_hist_900.tex};
            \addlegendentry{L=30}
        \addplot[blue] table [mark=none, x=e, y=pe, col sep=space] {Plots/PA_hist_1600.tex};
        	\addlegendentry{L=40}
        \addplot[green] table [mark=none, color = green, x=e, y=pe, col sep=space] {Plots/PA_hist_2500.tex};
        	\addlegendentry{L=50}
        \addplot[black] table [mark=none, x=e, y=pe, col sep=space] {Plots/PA_hist_3600.tex};
        	\addlegendentry{L=60}
        \end{axis}
    \end{tikzpicture}
    \caption{Distribution $\rho_\beta(e)$ of the energy per spin, $e$ at the transition temperature calculated with population annealing for system sizes $L =$ 30, 40, 50, and 60 (from top to bottom).}
    \label{fig:canonical_histogram}
\end{figure}
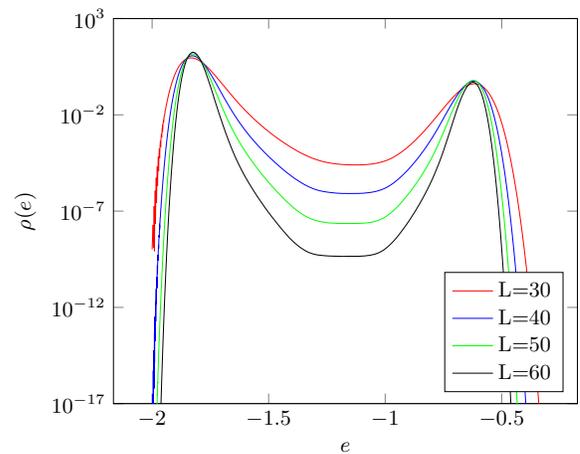

\section{\label{sec:Results}Results}
In this section we describe the results of our simulations.  Our primary objective is comparing the three annealing algorithms but we also obtain precise finite-size results for the Potts model observables described above.

In order to obtain results in the canonical ensemble at the transition temperature, the  microcanonical entropy is measured at every energy using each of the microcanonical annealing algorithms and the results are reweighted to obtain the free energy at the transition using Eqs.\ \eqref{eq:Z} and \eqref{eq:F}. The  peak ratio observables, $\rc$ and $\A$ are measured using Eqs.\ \eqref{eq:peakObs} and \eqref{eq:disordered_excess}, respectively. Ordered and disordered phase energies per spin, $e_\ord$ and $e_\dis$, are calculated from the canonical energy histogram, Eq.\ \eqref{eq:canonical_histogram} using Eqs.\ \eqref{eq:peakObs}.  For the system sizes considered, the ordered and disordered peaks in the energy distribution at coexistence, shown in Fig.\ \ref{fig:canonical_histogram}, are sufficiently separated that any ambiguity in defining breakpoint energy, $E_\od$ between ordered and disordered phases is negligible, and in practice $E_\od$ was chosen to be the point of minimum weight.  The reported values of all observables are obtained from a weighted average over $M$ independent simulations as described in Sec.\ \ref{sec:weight} and error bars are obtained by bootstrapping over these runs.  

Table  \ref{tab:results} shows results from the three algorithms and four system sizes for $e_\dis$, $e_\ord$, $\rc$, $\A$ and $\mathrm{var}(\beta F)$.   The results for SA are accurate to four or five significant digits and the results for the different algorithms are in agreement with one another within error bars.  The SA results for size $L=40$ are consistent with and more accurate than the results reported in Ref.\ \cite{JaKa97}.

\begin{figure}[h]
    \begin{tikzpicture}
        \pgfplotsset{   compat=newest,
                        scaled ticks=false,
                        yticklabel style={
                                /pgf/number format/precision=3
                        },
                    }
        \begin{axis}[   xlabel= {$\beta f$},
                        ylabel={$e_\dis$},
                        scatter/classes={
		                    SA={blue},
		                    HA100={green},
		                    PA={red}
		                },
            		    /pgf/number format/precision =4
            		]
            \addplot[   scatter,only marks, mark size = 1.0pt,%
                        scatter src=explicit symbolic
                    ]
            table [meta =alg, x=bf, y=ed, col sep=space] {Plots/bf_scatters_1600.txt};
            \legend{$SA$,$HA$,$PA$}
        \end{axis}
    \end{tikzpicture}
    \caption{Scatter plot of the dimensionless free energy per spin, $\beta f$ and the disordered phase energy per spin $e_\dis$ for each of $M=40$ runs for SA (blue), HA (green), and PA (red) at the transition temperature for system size $L=40$.}
    \label{fig:bf_ed_scatter}
\end{figure}
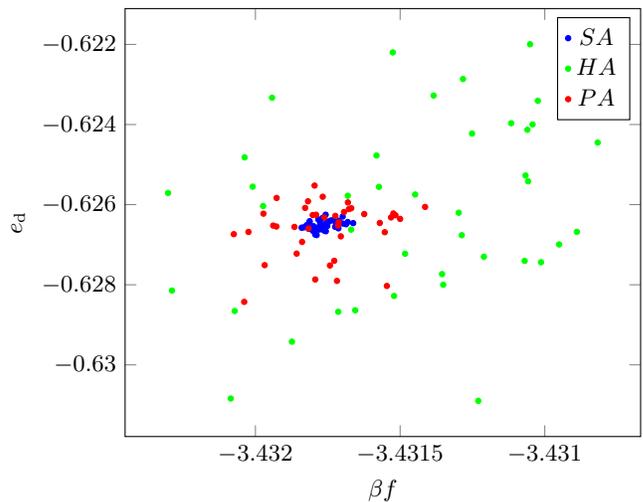

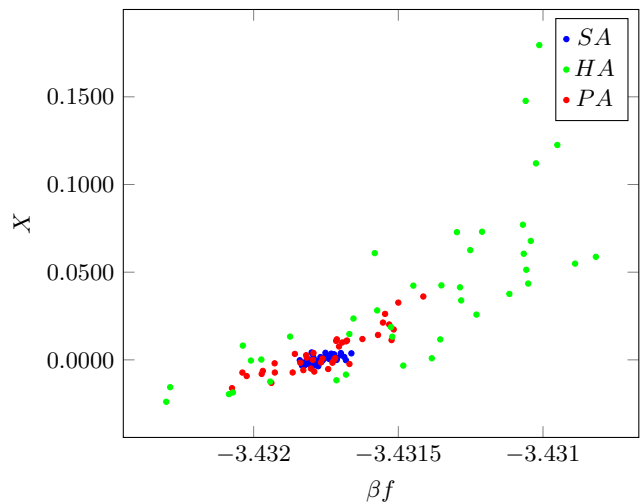
\begin{figure}
    \begin{tikzpicture}
        \pgfplotsset{compat=newest,
                     scaled ticks=false,
                     yticklabel style={
                        /pgf/number format/.cd,
                        fixed,
                        fixed zerofill,
                        /tikz/.cd
                        },
                    }
        \begin{axis}[xlabel= {$\beta f$}, ylabel={$\A$},scatter/classes={%
                		SA={blue},%
                		HA100={green},%
                		PA={red}},%
                		/pgf/number format/precision = 4
                		]
                \addplot[scatter,only marks, mark size = 1.0pt,%
                scatter src=explicit symbolic]%
                table [meta=alg, x=bf, y=A, col sep=space]{Plots/bf_scatters_1600.txt};
            \legend{$SA$,$HA$,$PA$}
        \end{axis}
    \end{tikzpicture}
    \caption{Scatter plot of the dimensionless free energy per spin, $\beta f$ and disordered phase excess, $\A$ for each of $M=40$ runs for SA (blue), HA (green), and PA (red) at the transition temperature for $L=40$.  The exact value in thermodynamic limit, $X=0$, corresponds to peak ratio $\rc=q$.}
    \label{fig:bf_X_scatter}
\end{figure}

Based on both the error bars, which quantify statistical errors, and on $\mathrm{var}(\beta F)$, which quantifies systematic errors (see Eq.\ \eqref{eq:syserrorcan}, we see that SA yields significantly more accurate and well-equilibrated results than either PA or HA.  Furthermore, except for size $L=60$, PA is better than HA. The relative ranking of the three algorithms is also visible in Figs.\ \ref{fig:bf_ed_scatter} and \ref{fig:bf_X_scatter}, which are scatter plots of $e_\dis$ and $\A$, respectively, vs.\ the free energy per spin, $\beta f$, for size $L=40$.  Each point represents one of the $M$ independent runs.  The larger the variance of the value of the observable, the larger statistical errors for the observable while a large covariance with $\beta f$ indicates a large systematic error according to \eqref{eq:syserrorcan}.

While measuring observables at the transition can be done relatively efficiently using the microcanonical ensemble, the simulations become more difficult with increasing system size.  It is believed that the major sources of hardness in these simulations are the four microcanonical phase transitions described in Sec.\ \ref{sec:model}.  To investigate this we used the integrated autocorrelation time of the magnetization $\tau$ as a measure of computational hardness.  

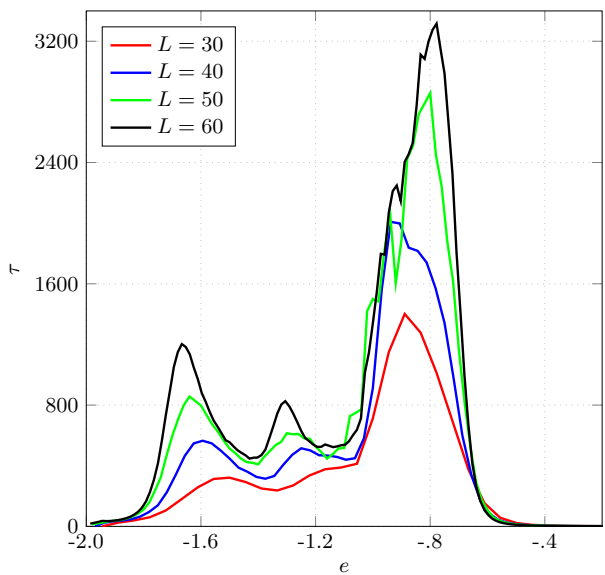
\begin{figure}
    \hspace{-12pt}
    \begin{tikzpicture}[yscale=.9, xscale = .9]
            \begin{axis}[%
                grid=major, grid style = {dotted},
                width=3in,
                height=3in,
                scale only axis,
                xmin=-2,
                xmax=-0.2,
                every y tick label/.append style={font=\color{black}},
                ymin=0,
                ymax=3400,
                ytick       ={    0, 800,  1600, 2400,  3200},
                ylabel      ={$\tau$},
                xlabel      ={$e$},
                yticklabels ={    0, 800,  1600, 2400,  3200},
                xtick       ={-.4,-.8,-1.2, -1.6, -2.0},
                xticklabels ={-.4,-.8,-1.2,-1.6,-2.0},
                legend pos= north west
                ]
            \addplot [
                color=red,
                solid,
                line width=1.0pt,
                ] table[mark=none, x=e, y=T, col sep=space]{Plots/autocorrelation_30.txt};
                \addlegendentry{$L=30$}
            \addplot [
                color=blue,
                solid,
                line width=1.0pt,
                ] table[mark=none, x=e, y=T, col sep=space]{Plots/autocorrelation_40.txt};
                \addlegendentry{$L=40$}
            \addplot [
                color=green,
                solid,
                line width=1.0pt,
                ] table[mark=none, x=e, y=T, col sep=space]{Plots/autocorrelation_50.txt};
                \addlegendentry{$L=50$}
            \addplot [
                color=black,
                solid,
                line width=1.0pt,
                ] table[mark=none, x=e, y=T, col sep=space]{Plots/autocorrelation_60.txt};
                \addlegendentry{$L=60$}
        \end{axis}
    \end{tikzpicture}
    \caption{Integrated autocorrelation time $\tau$ in units of Monte Carlo sweeps, as a function of energy ceiling per spin $e$ for system sizes $L=30$, 40, 50 and 60.}
    \label{fig:autocorrelation_by_size}
\end{figure}

We plot $\tau$ in Fig.\ \ref{fig:autocorrelation_by_size} as function of the energy ceiling for the four system sizes considered.  Indeed, $\tau$ displays sharp peaks that grow with system size.  While only three peaks are visible for the system sizes considered, we believe that the largest, rightmost peak is associated with both the condensation transition and the first wrapping transition.  For larger system sizes we expect that this peak would break into two distinct peaks.

\begin{figure}
    \hspace{-12pt}
    \begin{tikzpicture}[yscale=.9, xscale = .9]
        \begin{axis}[
            grid=major, grid style = {dotted},
            xmode=log,
            ymode=log,
            xlabel={cluster size},
            ylabel={$\csh$},
            ]
        \addplot [
            color=red,
            solid,
            line width=1.0pt,
            ] table[mark=none, x=size, y=csh, col sep=space]{Plots/cluster_size_histogram_0.txt};
            \addlegendentry{$e=-1.56$}
    
        \addplot [
            color=blue,
            solid,
            line width=1.0pt,
            ] table[mark=none, x=size, y=csh, col sep=space]{Plots/cluster_size_histogram_1.txt};
            \addlegendentry{$e=-1.73$}
            
        \addplot [
            color=green,
            solid,
            line width=1.0pt,
            ] table[mark=none, x=size, y=csh, col sep=space]{Plots/cluster_size_histogram_2.txt};
            \addlegendentry{$e=-1.84$}
         \addplot [
            color=red,
            dashed,
            line width=1.0pt,
            ] table[mark=none, dashed, x=size, y=fit, col sep=space]{Plots/cluster_size_histogram_0.txt};
            
        \addplot [
            color=blue,
            dashed,
            line width=1.0pt,
            ] table[mark=none, dashed, x=size, y=fit, col sep=space]{Plots/cluster_size_histogram_1.txt};
    
        \end{axis}
    \end{tikzpicture}
        \caption{Disordered cluster size histograms, $\csh$ for ceiling energies per spin $e=-1.56$ (red),  $-1.73$ (blue), and $-1.84$ (green) showing the behavior of $\csh$ above, near, and below the evaporation transition energy $e_4$, respectively, for size system $L=40$.  Dashed lines represent fitted power laws to the histograms for cluster sizes below a cutoff of 35.}
        \label{fig:csh}
\end{figure}
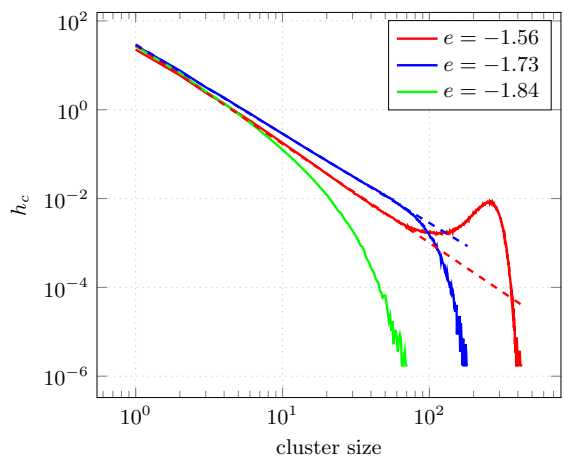

The evaporation transition is studied using methods described in Sec.\ \ref{sec:observables}.  The disordered cluster size histogram $\csh$ is plotted in Fig.\ \ref{fig:csh} at three energies for size $L=40$.  As described in Sec.\ \ref{sec:observables} we observe a power law with an additional distinct peak for energies above the evaporation transition,  power law behavior at the transition, and exponential decay below.  The integral of the peak above the fitted power law in Fig.\ \ref{fig:csh} gives a measurement of the  $\omega_c$. The power law behavior of $\csh$ at the evaporation transition is unexpected and warrants further investigation.

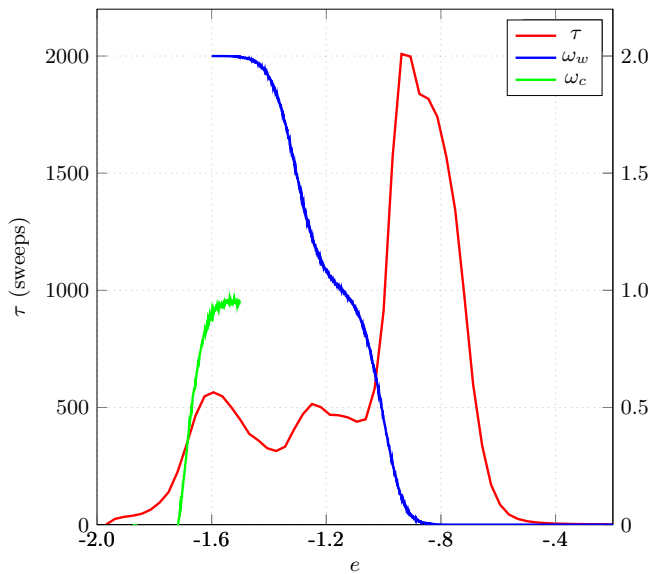
\begin{figure}[h]
    \hspace{-12pt}
    \begin{tikzpicture}[yscale=.9, xscale = .9]
        \begin{axis}[%
            grid=major, grid style = {dotted},
            width=3in,
            height=3in,
            scale only axis,
            xmin=-2,
            xmax=-0.2,
            every y tick label/.append style={font=\color{black}},
            ymin=0,
            ymax=2200,
            ytick       ={    0, 500,  1000, 1500,  2000},
            ylabel      ={$\tau\ (\textrm{sweeps})$},
            xlabel      ={$e$},
            yticklabels ={    0, 500,  1000, 1500,  2000},
            xtick       ={-.4,-.8,-1.2, -1.6, -2.0},
            xticklabels ={-.4,-.8,-1.2,-1.6,-2.0},
            ]
        \addplot [
            color=red,
            solid,
            line width=1.0pt,
            ] table[mark=none, x=e, y=T, col sep=space]{Plots/autocorrelation_time.tex};
            \addlegendentry{$\tau$}
            \label{autocorrelation}
        \end{axis}
        \begin{axis}[%
            width=3in,
            height=3in,
            scale only axis,
            axis x line*=none,
            axis y line*=right,
            ylabel={$\omega$},
            ylabel near ticks, yticklabel pos=right,
            every y tick label/.append style={font=\color{black}},
            ymin=0,
            ymax=2.2,
            xmin=-2.0,
            xmax=-0.2,
            ytick       ={    0, .5,  1.0, 1.5,  2.0},
            yticklabels ={    0, 0.5,  1.0, 1.5,  2.0},
            xtick       ={-.4,-.8,-1.2, -1.6, -2.0},
            xticklabels ={-.4,-.8,-1.2,-1.6,-2.0},
            ]
        \addlegendimage{/pgfplots/refstyle=autocorrelation}
        \addlegendentry{$\tau$}
        \addplot [
            color=blue,
            solid,
            line width=1.0pt,
            ] table[mark=none, x=e, y=w, col sep=space]{Plots/wrapping.tex};
            \addlegendentry{$\omega_w$}
        \addplot [
            color=green,
            solid,
            line width=1.0pt,
            ] table[mark=none, x=e, y=c, col sep=space]{Plots/condensation.tex};
            \addlegendentry{$\omega_c$}
        \end{axis}
    \end{tikzpicture}
    \caption{The integrated autocorrelation time $\tau$ (red, left axis) is plotted against ceiling energy per spin $e$ over the entire simulation energy range $-2 < e < 0$ for $L=40$. The wrapping fraction $\omega_w$ (blue, right axis) is plotted from $-1.6 < e < .7 $ and the \drop fraction $\omega_c$ (green, right axis) is plotted from $-1.85 < e < -1.5$.  The first two peaks in $\tau$, from right to left, coincide with jumps in $\omega_w$ that signify the first and second wrapping transitions, respectively.  The third peak in $\tau$ coincides with a rapid drop of $\omega_c$ signifying the evaporation transition.  
        }
    \label{fig:autocorrelation}
\end{figure}

In Fig.\ \ref{fig:autocorrelation} we show the magnetization integrated autocorrelation time, $\tau$, the wrapping fraction, $\omega_w$, and \drop fraction, $\omega_c$, for $L=40$.  The wrapping fraction  is zero at high energies and displays two jumps as expected in the coexistence region $e_o < e < e_d$.  These jumps coincide with the two rightmost peaks in $\tau$, suggesting that they are in fact points where increased computational effort is needed. 

We observe that $\omega_c \rightarrow 1$ for energies far above the condensation transition, and drops quickly to zero below the transition at energy $e_4$.  The point where $\omega_c$ begins to rapidly decrease marks the point where a droplet of disordered phase becomes unstable, which coincides in Fig.\ \ref{fig:autocorrelation} with the final peak in $\tau$. 

The growth of the the autocorrelation time due to the wrapping and condensation/evaporation transitions poses the greatest challenge to any microcanonical algorithm including microcanonical annealing.  These transitions limit the system sizes that can be feasibly studied and ultimately will lead to exponential growth in the resources required to perform simulations.  However, in Fig.\ \ref{fig:autocorrelation_by_size} we observe only a modest increase in $\tau$ with increasing system size.  Compare this to the canonical energy histogram in Fig.\ \ref{fig:canonical_histogram}, where the depth of the valley between ordered and disordered grows exponentially for the same sizes considered.  The depth of this valley provides an estimate for the time scale needed by a canonical MCMC algorithm to tunnel between an ordered phase and the coexisting disordered phase.  This suggests that for modest system sizes microcanonical annealing is not only more efficient than canonical MCMC but also achieves better scaling until exponential slowing eventually dominates.     

\begin{center}
	\begin{table*}
		\begin{tabular}{ |c|c|c|c|c|c|c|} 
			\hline
			$L$ & Algorithm & $e_\dis$ & $e_\ord$ & $\rc$& $\A$ &$\mathrm{var}(\beta F)$\\
			\hline
            \multirow{4}{*}{30} & 		   SA		    & -0.626551(25) &-1.820723(12)&19.965(97)&7.9e-5 $\pm$ 2.2e-4 &0.0013\\
                                &  		   HA 		    & -0.62678(73) &-1.81966(90)&18.1(1.6)&4.9e-3 $\pm$ 8.1e-3  &0.38\\ 
                                &  		   PA		    & -0.626603(48) &-1.820732(49)&20.17(29)&-3.9e-4 $\pm$ 6.6e-4&0.012\\
            \hline
            \multirow{4}{*}{40} & 		   SA		    & -0.626514(34)&-1.820632(17)&19.84(15)&3.6e-4 $\pm$  3.5e-4&0.0049\\
                                &  		   HA 		    & -0.62619(49)&-1.82017(32)&15.0(1.0)&1.5e-2 $\pm$ 7.1e-3 &0.41\\ 
                                &  		   PA		    & -0.62661(11)&1.82050(11)&19.39(69)&1.4e-3 $\pm$ 1.8e-3  &0.064\\
            \hline
            \multirow{4}{*}{50} & 		  SA		    & -0.626525(41)&-1.820649(20)&19.45(31)&1.3e-3 $\pm$ 7.6e-4 &0.072\\  		
                                &  		   HA 		    & -0.62689(19)&-1.81991(30)&15.1(1.1)&1.4e-2 $\pm$ 7.8e-3&0.44\\
                                &  		   PA		    & -0.62663(12)&-1.82072(15)&21.5(1.2)&-3.5e-4 $\pm$ 4.2e-3&0.25\\
            \hline
            \multirow{4}{*}{60} & 		   SA		    & -0.626537(41)&-1.820753(31)&19.62(46)&8.7e-4  $\pm$ 1.1e-3 &0.096\\   					
                                &  		   HA 		    & -0.62652(18)&-1.82050(36)&20.31(91)&-7.0e-4 $\pm$ 5.8e-3  &0.70\\ 
                                &  		   PA	         &-0.62660(18)&	-1.82081(34)& 22.93(83)&-5.8e-3 $\pm$ 1.7e-2& 1.3\\
            \hline
            $\infty$ (exact)				&			-			&-.626529\ldots &-1.820684\ldots&20& 0 &	\\
            \hline
		\end{tabular}
		\caption{Energies of the disordered phase, $e_\dis$,  ordered phase, $e_\ord$,  the peak ratio, $\rc$, disordered phase excess $\A$, and the variance of the free energy, $\mathrm{var}(\beta F)$, at the transition temperature.  Results are shown for the SA, HA and PA simulations, and exact values in the thermodynamic limit \cite{Ba73}.}
	\label{tab:results}
	\end{table*}
\end{center}

\section{Discussion}
\label{sec:discussion}
We have introduced a class of equilibrium annealing algorithms that includes population annealing and an equilibrium version of simulated annealing.  We have implemented algorithms in this class in the microcanonical ensemble and applied them to the 20-state Potts model, which displays a strong first-order transition.  In agreement with previous work, we find that simulating a thermal first-order transition can be effectively carried out in the microcanonical ensemble yielding high precision results. In these applications of microcanonical annealing, the purpose of annealing is to collect data at all energies in the coexistence region in order to obtain canonical averages at the phase transition temperature using reweighting.

We compared the performance of equilibrium simulated annealing, population annealing and a hybrid algorithm that interpolates between the two.  For the system sizes and number of sweeps considered here, equilibrium simulated annealing was found to be the most efficient algorithm followed by population annealing.  The hybrid algorithm performed worst except for the largest size ($L=60$) where it slightly outperformed population annealing, though for this size neither algorithm equilibrated the system with the allotted number of Monte Carlo sweeps.  

We conjecture that SA outperforms PA and HA because it takes advantage of the exponential convergence to equilibrium of Markov chain Monte Carlo in contrast to the $1/R$ convergence to equilibrium of sequential Monte Carlo.  Note that for $L=40$ the magnetization integrated autocorrelation time $\tau$ is bounded by  $\tau \leq 2000$ while the number of sweeps per annealing step in this region is $7.6 \times 10^6$ so there are more than  $10^3\tau$ sweeps at every energy and, assuming the rate of convergence to equilibrium is approximately $\tau$, systematic errors will be extremely small. By contrast, in PA the number of sweeps per annealing step carried out in the transition region is much less than $\tau$ and equilibration is achieved primarily by resampling, which converges only as $1/R$.  The hybrid algorithm suffers from both an inadequate population size for effective resampling and too few Monte Carlo sweeps.  

A toy model illustrates the above intuition. Suppose that the distance from equilibrium, $\Delta$ behaves as $\Delta = (1/R)e^{-t/\tau}$ where $R$ is the population size, $t$ the number of MCMC sweeps for each replica at each temperature and $\tau$ is the autocorrelation time of the MCMC, which is here assumed to be constant.  We allot each algorithm the same total number of sweeps per annealing step so that the scaled number sweeps, $c$, defined as $c=Rt/\tau$ is held fixed. Note that $c$ is the number of autocorrelation times for each annealing step in simulated annealing.  In terms of $R$ and $c$ we seek to minimize $(1/R)e^{-c/R}$ for fixed $c$ subject to the constraints that the population size is at least one, $R \geq 1$, and that at least one MCMC sweep is performed on each replica at each annealing step, $t \geq 1$.  The latter constraint is required to insure that some exploration and decorrelation is carried out at each annealing step. In terms of population size, this constraint can be written as $R \leq c\tau$. It is straightforward to show that there are two minima for $\Delta$ and they occur at the two endpoints,  SA ($R=1$), and  PA ($R = c\tau$). At the SA endpoint $\Delta = e^{-c}$ while at the PA endpoint $\Delta \approx 1/c\tau$.  In addition there is a maximum at $R=c$ where $\Delta = (1/ec)$.  While this toy model cannot be expected to provide quantitative results, it is expected to yield qualitative comparisons between algorithms.

The first conclusion from the toy model is that if the autocorrelation time is sufficiently short that $t/\tau$ can be made large, then it is better to use SA.
A second conclusion is that it does not pay to compromise between many MCMC sweeps and a large population.  In our case $c$, at least in the difficult coexistence region, is approximately $10^3$ so that $\Delta_{\rm SA} \ll \Delta_{\rm PA}$ and we expect that SA is the preferred algorithm.   The hybrid algorithm with $R=10^2$ is not far from the worst choice, $R=c$ while PA is closer to the second optimum at $R=c \tau$ so we expect that for our choices of parameters HA would perform worse that PA. 

We can conclude from both the simulation results and the simple toy model that SA is the best of the microcanonical annealing methods for high precision studies of the transition in large-$q$ Potts models and, likely, other systems with thermal first-order transitions and simple symmetry breaking.  
On the other hand, for systems that display a rough free energy landscape without obvious symmetries, we expect PA or HA to outperform SA.  In each run of SA for the $q$-state Potts model only one of the $q$ low temperature phases is discovered because the barriers between phases are too high.  This deficiency does not affect any of the observables we measured because of the symmetry between the $q$ low temperature phases.  The situation is different in the case of multiple inequivalent phases or a glassy phase with a rough free energy landscape.   In order to apply SA to a system with a free energy landscape with multiple inequivalent minima separated by high barriers it is necessary to combine many runs of SA using weighted averaging. Good statistics requires a large number of runs so that each important minimum is found in many runs.   We conjecture that it is better to do a single run of HA or PA with the same annealing schedule as for SA and with $R=M$ replicas where $M$ is the number of runs used for SA. For HA or PA, the resampling step distributes the work assigned to each minimum according to the weight of that minimum so that the more important minima are more accurately sampled and minima that are irrelevant at the target low temperature are removed from the population.  Evidence supporting this conjecture can be found in Ref.\ \cite{WaMaKa15}, where it was shown that PA with population size $R$ is far more likely to find ground states of Ising spin glasses than $M=R$ runs of SA with the same annealing schedule. 

The microcanonical annealing algorithms discussed here may also be useful for sampling equilibrium systems with rough free energy landscapes and solving hard combinatorial optimization problems in situations where behavior similar to first-order transitions occurs.  For example, spin glasses display temperature chaos \cite{McBeKi82,BrMo87,WaMaKa15a}, which is the phenomena that typical states of the system changes discontinuously with temperature.  It would be interesting to see whether microcanonical population annealing performs better than canonical population annealing for systems with temperature chaos.

\acknowledgements
This work was supported in part by the National Science Foundation (Grant No.~DMR-1507506).  We thank Chris Amey, Gili Rosenberg, Martin Weigel, Roman Koteck{\'{y}}, and Helmut Katzgraber for useful discussions.

\bibliographystyle{unsrt}
\bibliography{bib}

\end{document}